\documentstyle[11pt,paspconf]{article}

\begin{document}

\title{High-$z$ radio galaxies and the `Youth-Redshift Degeneracy'}
\author{Katherine M.\ Blundell \& Steve Rawlings}
\affil{Oxford University Astrophysics, Keble Road, Oxford, OX1
3RH, UK}

\begin{abstract}
We discuss a unifying explanation for many `trends with redshift'
of radio galaxies which includes the relevance of their ages
(time since their jet triggering event), and the marked
dependence of their ages {\em on redshift} due to the selection
effect of imposing a flux-limit.  We briefly describe some
important benefits which this `youth-redshift degeneracy' brings.
\end{abstract}


\section{Trends in COSMIC EPOCH or trends in SOURCE AGE?}

With radio source ages of at most a few $10^8$ years, the huge
decrease in co-moving space density of luminous radio-sources
from redshift $z = 2$ to $z = 0$ (Longair 1966, Dunlop \& Peacock
1990) is not due to a decline in luminosity of individual
objects.  However, it does not follow that one can ignore the
luminosity evolution of the individual radio-sources and their
ages in all studies of their `cosmic evolution'.

Application of a flux-limit in any model of radio-galaxy
evolution in which the luminosity $(P)$ decreases with time means
that all observable high-redshift radio-galaxies must be seen
when the lobes are young ($< 10^7$ years old; see Figure 1).
This mechanism (for details see Blundell \& Rawlings 1999) is
responsible for the highest-redshift members of any low-frequency
survey of radio-sources (such as 3C) having significantly more
powerful jets and being significantly younger than the more local
members.  This physical mechanism plays a crucial role in
explaining `trends with redshift', without invoking any intrinsic
or strong environmental differences between the radio-galaxies
seen at low-$z$ and high-$z$; we discuss five such trends below.

\begin{figure}[!t]
\plotfiddle{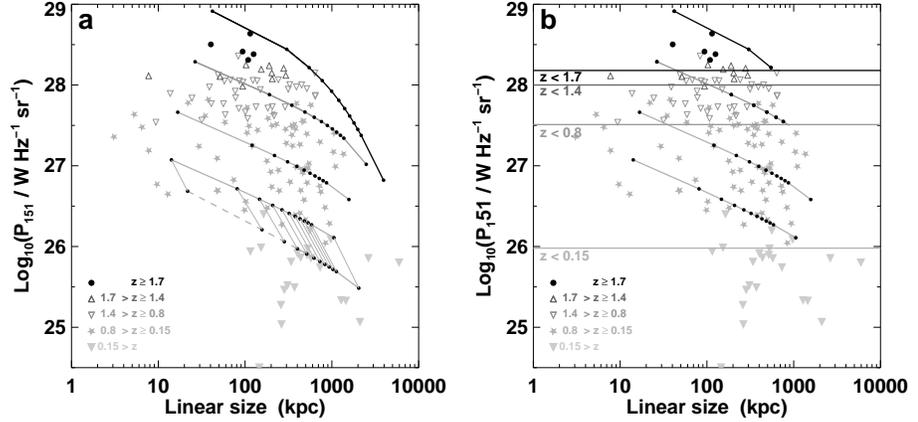}{6cm}{0}{70}{70}{-200}{-370}
\caption{Overlaid on the `$P$--$D$' plane for the 3C
sample in {\em \bf a} are model tracks tracing out the evolution
of four example radio sources in luminosity and linear size with,
from top to bottom, $Q = 5 \times 10^{39}$ W at $z = 2$, $Q = 1
\times 10^{39}$ W at $z = 0.8$, $Q = 2 \times 10^{38}$ W at $z =
0.5$ and $Q = 5 \times 10^{37}$ W at $z = 0.15$.  The dashed line
indicates how the lower track luminosity reduces by $<$ half an
order of magnitude if the ambient density becomes an order of
magnitude lower.  In {\em \bf b} the horizontal lines represent
the luminosities at which the flux-limit of 12 Jy takes its
effect at the different redshifts indicated.  A combination of
the dramatically declining luminosity-with-age of the high-$Q$
sources, their scarcity in the local Universe, together with the
harsh reality of the flux-limit means that very powerful sources
with large linear sizes are rarely seen.}
\end{figure}

(1) The linear size evolution which is observed in low-frequency
flux-limited samples of classical double radio-sources (Kapahi et
al.\ 1987, Barthel \& Miley 1988, Blundell, Rawlings \& Willott
1999) arises because the high--$z$ sources are younger, hence tend
to be shorter.  Falle (1991) has shown that higher jet-powers
($Q$) increase the rate at which radio-source lengths ($D$) grow
with time $t$.  This positive dependence on jet-power of the rate
at which the lobe-lengths grow contributes to the linear size
evolution being as mild as it is (Blundell, Rawlings \& Willott
1999).

(2) Barthel and Miley (1988) had suggested that higher redshift
environments are denser and more inhomogeneous than at low
redshift since they found increased distortion in the structures
of their high-$z$ sample of steep-spectrum quasars compared with
their low-$z$ sample.  Young radio sources which have recently
undergone a jet-triggering event [assumed to be a galaxy-galaxy
merger (Sanders et al.\ 1988)] may have the passage of their jets
considerably more disrupted where there is a higher density and
greater inhomogeneity in the ambient post-recent-merger
environment.  A general trend of denser inter-galactic
environments at high-$z$ cannot be inferred from their result.

(3) Where the alignment effect is caused by star-formation, it
will be more easily triggered close in to the host galaxy or
within the product of a recent merger than at distances further
out sampled by the head of an expanding radio-source later in its
lifetime.  Where the alignment effect is caused by dust-scattered
quasar light, the certain youthfulness of distant radio-galaxies
alleviates the near discrepancy (De Young 1998) between
radio-source ages and the time-scale for which dust grains can
survive in the presence of shocks caused by the advancing
radio-jets.  The `youth--redshift (YZ) degeneracy' is consistent
with Best et al's (1996) finding that the smallest sources in a
sample of $z \sim 1$ radio-galaxies (all with very similar
luminosities) are those which are most aligned with optical
emission.  Indeed Best et al.\ suggest this as the explanation of
their observation.

(4) Garrington \& Conway (1991) have found a tendency for
depolarisation to be higher in sources with higher $z$.  Objects
with higher $z$ which are younger will be in much more recently
merged environments with the consequence that inhomogeneities in
density or magnetic field will more readily depolarise the
synchrotron radiation from the lobes.  Moreover, higher-$z$
sources being younger and somewhat shorter will be closer in to
the centre of the potential well.  The higher density in this
region will enhance the observed depolarisation.

(5) Many of the highest--$z$ radio-galaxies have gas masses
comparable to gas-rich spiral galaxies (Dunlop et al 1994, Hughes
et al 1998) and inferred star-formation rates which, in the local
Universe, are rivalled only by galaxy-galaxy mergers like Arp 220
(Genzel et al 1998).  If high-$z$ objects are being viewed during
a similar merging of sub-components, the associated star
formation could be responsible for a significant fraction of the
stellar mass in the remnant galaxy.  Since the high-$z$
radio-galaxies, those which have been detected by Hughes et al
(1988) using SCUBA, are necessarily young ($< 10^7$ years, see
Fig.\ 1), and since the whole merger must take a few dynamical
crossing times, or $10^{8-9}$ years, the implication is that the
event which triggered the jet-producing central engine is
synchronised with massive star formation in a gas-rich system,
perhaps as material streams towards the minimum of the
gravitational potential well of the merging system.

The YZ degeneracy may help explain why few lower-$z$
radio-galaxies show similarly large (rest-frame) far-infrared
luminosities compared to the high-$z$ population: they are being
observed significantly longer after the jet-triggering event.

\section{The elusiveness of cosmic parameters}

A variant on Fig.\ 1 also includes the location on the $P$--$D$
plane of the most extreme redshift ($z > 3$) radio galaxies
known.  Such a figure may be found in our recent letter to {\em
Nature} (Blundell \& Rawlings 1999).  When these are plotted for
different cosmological models, significant though subtle
differences emerge for the high-$z$ sources which illustrate the
difficulty of distinguishing between different underlying cosmic
geometries when more dramatic influences such as the
YZ-degeneracy, and variations in source environments, are at
work.  In rough order of importance to the distribution of
sources on the $P$--$D$ plane, we have:
\begin{enumerate}
\item What is the finding frequency of the
survey in the rest-frame of the source?  (see Blundell, Rawlings
\& Willott [{\sl astro-ph/9907418}])
\item What is the flux-limit? This excludes
faint/old objects at high redshift.
\item What is Q, the jet-power?
\item What is the ambient density into which
the radio-source is expanding?
\item What density profile is the
radio-source expanding into?
\item What is the cosmic geometry?
\end{enumerate}
The use of double radio sources as `standardizable'
rods (e.g.\ Daly 1994) is beyond reach.

\section{The benefits of the YZ degeneracy }

Since extreme-$z$ radio galaxies are young, all with similar
$Q$, they deliver the fine time-resolution required for the
solution of problems which it may be difficult to study with
objects like optically-selected quasars, whose ages are
indeterminate: examination of the environments of distant radio
galaxies provides a snapshot of the host galaxy evolutionary
status just after the jet-triggering event.

At redshift $\sim 4$ we observe radio galaxies $\sim 1$ \,Gyr
after the Big Bang and in environments which saw a jet-triggering
event no earlier than $10^7$ years prior to that.  This time-step
is over an order of magnitude smaller than the dynamical crossing
time of a massive galaxy, and 2 orders of magnitude smaller than
the age of the Universe at the epochs probed, giving fine
time-resolution essential to any study of triggering (and hence
merging) rates at early cosmic epochs.

The YZ-degeneracy is unavoidable, and implies a wide and
increasingly ill-defined gulf between the `luminosity function'
and the `birth function' of radio-galaxies.  The luminosity
function is a super-set of the radio-sources which make it above
the various flux-limits.  The `birth function' measures the
trigger rates of radio galaxies.  We have carefully developed a
model for radio-source evolution which for the first time in
radio-source modelling incorporates the role played by the
hotspots.  This, together with a proper treatment for the
interception of radio galaxies born at successively lower $z$ with
our light-cone, enables us to perform Monte-Carlo simulations to
constrain the birth-function of radio galaxies out to high-$z$
(Blundell, Rawlings \& Willott, in prep).

\end{document}